\documentclass[12pt]{article}
\usepackage{amsmath}
\usepackage{amssymb}
\usepackage{graphicx}
\usepackage[auth-lg,affil-sl]{authblk}

\newcommand\R{\mathbb{R}}
\newcommand\vol{\mathrm{vol}}
\newcommand\E{\mathsf{E}}
\newcommand\Z{\mathbb{Z}}
\newcommand\N{\mathbb{N}}
\usepackage{empheq}
\usepackage{subfigure}

\usepackage{color}

\title{Relative entropy as a measure of inhomogeneity in general relativity}

\begin{document}

\author{Nikolas Akerblom\footnote{{\tt nikolasa@nikhef.nl}}}
\affil{Nikhef Theory Group\\Amsterdam, The Netherlands}

\author{Gunther Cornelissen\footnote{{\tt g.cornelissen@uu.nl}}}
\affil{Department of Mathematics\\Utrecht University, The Netherlands}

\maketitle

{\abstract {\noindent
We introduce the notion of relative volume entropy for two spacetimes with preferred compact spacelike foliations. This is accomplished by applying the notion of Kullback-Leibler divergence to the volume elements induced on  spacelike slices. The resulting quantity gives a lower bound on the number of bits which are necessary to describe one metric given the other. For illustration, we study some examples, in particular gravitational waves, and conclude that the relative volume entropy is a suitable device for quantitative comparison of the inhomogeneity of two spacetimes.

\begin{flushright} NIKHEF 2010-025 \end{flushright}}}

\newpage

\section{Introduction}
How much information is required to describe one spacetime in terms of another? More detailed manifestations of this question are: Just how much more complex is a solution of the Einstein equation in the presence of matter than a vacuum solution? How ``complicated'' is a gravitational wave? In this paper, we provide a possible answer to these questions in terms of the relative entropy of the volume elements associated with the metric tensors of two given spacetimes.

Our approach is inspired by two well-known facts:
\begin{itemize}
\item In mathematical information theory, relative information is measured by the Kullback-Leibler divergence, or relative entropy, of probability measures \cite{KL}.
\item In order for our answer to make sense, we want it to be a diffeomorphism invariant of the ordered pair of metrics.
A natural covariant of a pseudo-Riemannian metric is the Radon-Nikodym derivative of its associated measure on spacelike slices.\footnote{Recall that the Radon-Nikodym derivative $d\nu/d\mu$ of two measures $\nu$ and $\mu$, with $\nu$ absolutely continuous w.r.t.\ $\mu$, is a function $\phi$ such that $\!\int f\, d\nu = \int\!f \cdot \phi\, d\mu$. Here, $\phi$ is unique up to a set of $\mu$-measure zero, compare \cite{HS}, Ch.\ 19. Intuitively, we can think of the Radon-Nikodym derivative as a Jacobian for measures.}
\end{itemize}
We therefore propose to compute the relative entropy of two continuous probability distributions associated with our metrics, namely the normalized volume densities in spacelike slices. 

For this method to make sense without further complications, we restrict ourselves to the following situation. Suppose we are given a topological space
$$
M= \R_t\times X,
$$
where $X$ is compact, and two metrics
$$
g^{(k)}_{\mu\nu}dx^\mu dx^\nu=-dt^2+\gamma^{(k)}_{ij}dx^i dx^j\quad(k=1,2;\; i,j=1,\dots,d)
$$
on $M$. Define the normalized volume densities by
\begin{equation}\label{densdef}
\tilde{\rho}_k=\sqrt{\det\gamma^{(k)}_{ij}}/\vol_k,
\end{equation}
where
\begin{equation}\label{voldef}
\vol_k=\int d^dx\, \sqrt{\det\gamma^{(k)}_{ij}}.
\end{equation}

Then
$$
\tilde{\rho}_k\, d^dx
$$
is proportional to the physical volume in a (natural) spacelike slice corresponding to the coordinate volume $d^dx$.

Mathematically, the normalized volume densities can be interpreted as probability measures on $X$ (though we are still dealing with entirely deterministic physics). In a spacelike slice, they are absolutely continuous w.r.t.\ each other, and the Radon-Nikodym derivative of the measures $\tilde{\rho}_1\, d^dx$ and $\tilde{\rho}_2\, d^dx$ is equal to $\tilde{\rho}_1/\tilde{\rho}_2$.

We now introduce the Kullback-Leibler divergence (relative entropy) of our volume densities, the \emph{relative volume entropy}, as
$$
\E=\E(\tilde{\rho}_2\|\tilde{\rho}_1)=\int d^dx\, \tilde{\rho}_2 \log(\tilde{\rho}_2/\tilde{\rho}_1).
$$

The well-known information theoretic meaning of this entropy is that it measures the number of ``nats'' (or bits if we took the logarithm to the base $2$) that are necessary to describe the volume density $\tilde{\rho}_2$ when given $\tilde{\rho}_1$. 

For our problem of comparing spacetimes, this implies the following result: \emph{At least $\E \cdot \log_2{e}$ bits will be necessary to describe $(M,g^{(2)}_{\mu\nu})$ in terms of $(M,g^{(1)}_{\mu\nu})$.}

The relative entropy is known to be strictly positive for $\tilde{\rho}_1 \neq \tilde{\rho}_2$, and invariant under coordinate changes on spacelike slices.\footnote{It is a basic irritant in information theory that for continuous distributions the usual ``differential entropy'' $\int d^dx\, \tilde{\rho}_2 \log(\tilde{\rho}_2)$ \emph{does not} have these properties and one needs to fix a ``background distribution'' $\tilde{\rho}_1$. This is very much in line with Jaynes' viewpoint that entropy depends on a choice of reference frame \cite{Jaynes}.} For example, isometric metrics have vanishing relative entropy, since the volume elements transform by a trivial Jacobian.

Of course, our proposed information measure neglects some relative information by focussing solely on the volume densities. It appears to be a good measure for the difference in inhomogeneity between the metrics. 

To illustrate this last statement: A Kasner solution in any dimension $D>3$ has zero relative entropy over a Minkowski background, since its normalized volume density is independent of the parameters of the solution, cf.\ Section \ref{Kasner}. On the other hand, cosmological solutions in $2+1$-dimensions, whose volume densities have a ``lumpy,'' soliton-like behavior over a given vacuum background,\footnote{Similar spacetimes where studied by Deser-Jackiw-'t Hooft \cite{Deser:1983dr, Deser:1983tn}, and later in \cite{Gruzinov:2006nk}.} have a rather interesting relative entropy, cf.\ Section  \ref{Cosmo}. For example,  when space is spherical ($X=S^2$), and we place $n-1$ particles on the sphere, so that the resulting volume density is $\tilde{\rho}_n$, we obtain a relation of the form 
$$
\E(\tilde{\rho}_n\|\tilde{\rho}_0) \sim \log{n}\quad (n\to\infty).
$$
This certainly is a very satisfying result: \emph{The more particle sources are in the energy-momentum tensor, the more information is necessary to describe spacetime over a vacuum background.} By numerical integration, the very same result appears to hold true also for $X=T^2$, so we see that the relative entropy has little to do with cosmic topology, but rather with the amount of local fluctuation in volume.

We now proceed to work out some examples, in order to illustrate how the relative volume entropy behaves for different classes of spacetimes.

\section{Examples of relative volume entropies}
Here we study the relative volume entropy of Kasner metrics over (toroidal) Minkowski space (Sect.\ \ref{Kasner}), of cosmological spacetimes in $2+1$ dimensions with a positive cosmological constant (Sect.\ \ref{Cosmo}), and of exact gravitational waves in (toroidal) Minkowski space (Section \ref{Wave}).
\subsection{Kasner metrics \emph{vs.}\ Minkowski space}\label{Kasner}
Let
$$
M=\R_t\times T^d\quad(d>2),
$$
where $T^d$ is the $d$-torus.

The Kasner metric on this space is (see e.g.\ \cite{Misner:1974qy})
$$
ds^2_\text{Kasner}=-dt^2+\sum_{i=1}^d t^{2 p_i} (dx^i)^2,
$$
where the parameters $p_i$ satisfy
\begin{equation}\label{params}
\sum_{i=1}^d p_i =1,\quad \sum_{i=1}^d p_i^2 =1,
\end{equation}
and the coordinates on the torus are periodic with period-length $1$.

Since
$$
\gamma_{ij}^\text{(Kasner)}=t^{2p_i}\delta_{ij}\quad\text{(no sum on }i),
$$
using the first relation in \eqref{params}, we find that the volume \eqref{voldef} of Kasner is
$$
\vol_{\text{Kasner}}=t,
$$
so that the \emph{normalized} volume density \eqref{densdef} reads
$$
\tilde{\rho}_\text{Kasner}=1;
$$
just the same as for the Minkowski metric:
$$
\tilde{\rho}_\text{Kasner}=\tilde{\rho}_\text{Minkowski}.
$$
Consequently,
$$
\E(\tilde{\rho}_\text{Kasner}\|\tilde{\rho}_\text{Minkowski})=0.
$$
At face value, this result is hardly surprising, as the Kasner metric is just ``expanding Minkowski space.'' As mentioned in the Introduction, the relative volume entropy detects the difference in inhomogeneity between the metrics.

\subsection{Cosmology in 2+1 dimensions with $\Lambda>0$}\label{Cosmo}
A more interesting relative volume entropy arises for certain $2+1$-dimensional ``cosmological'' spacetimes with cosmological constant $\Lambda>0$ and pointlike matter.

Let's begin by reviewing the basic features of these spacetimes \cite{Deser:1983dr, Deser:1983tn,Gruzinov:2006nk}. The spacetime topology is
$$
M=\R_t\times X^2,
$$
with metric
$$
ds^2=-dt^2+a(t)^2\gamma_{ij}\,dx^i dx^j,
$$
where $X^2$ is a compact surface. We choose isothermal coordinates $x$, $y$ on $X^2$ and write the metric as 
\begin{equation}\label{metricansatz}
ds^2=-dt^2+a(t)^2e^\phi (dx^2+dy^2),
\end{equation}
and the Einstein equation as
$$
R_{\mu\nu}-\frac{1}{2}\,R\,g_{\mu\nu}+\Lambda\, g_{\mu\nu}=T_{\mu\nu},
$$
where, following \cite{Deser:1983dr}, we take $T_{\mu\nu}$ to be of the form appropriate for a cloud of non-interacting point particles $i$ of mass $m_i$ sitting at \emph{fixed} coordinates $\vec{x}_i=(x_i,y_i)$:
\begin{empheq}[left=\empheqlbrace]{equation*}
\begin{split}
T_{00}=a(t)^{-2}e^{-\phi}\sum_i m_i\delta^{(2)}(\vec{x}-\vec{x}_i),\\
T_{\mu\nu}=0\quad \text{for }(\mu,\nu)\neq(0,0).
\end{split}
\end{empheq}
By distributing the point particles in a suitable manner over the manifold $X^2$, one can obtain a matter distribution which satisfies the cosmological principles of homogeneity and isotropy.\footnote{Possibly modulo some subtle mathematical constraints. For the example of three particles, see \cite{Eremenko}.}

Plugging in our metric ansatz into the Einstein equation, we find that the only non-trivial components are\footnote{We define $\Delta=\partial_x^2+\partial_y^2$.}
\begin{empheq}[left=\empheqlbrace]{equation}\label{einsteq}
\begin{split}
\left(\frac{\dot{a}}{a}\right)^2-\frac{e^{-\phi}}{2a^2}\Delta\phi-\Lambda
&=a^{-2}e^{-\phi}\sum_i m_i\delta^{(2)}(\vec{x}-\vec{x}_i),
\\
-e^{\phi}a\ddot{a}+\Lambda a^2 e^\phi&=0.
\end{split}
\end{empheq}

The second equation in \eqref{einsteq} is readily integrated to give:\footnote{This result is at variance with \cite{Gruzinov:2006nk}, where the scale factor is polynomial in time, rather than exponential. This appears to be based on a slip in the integration in loc.\,cit.} 
$$
a(t)=c_1\cosh{\sqrt{\Lambda}\,t}+c_2\sinh{\sqrt{\Lambda}\,t}.
$$
For convenience, let us pick the solution where $c_1=1$ and $c_2=0$, such that
$$
a(t)=\cosh{\sqrt{\Lambda}\, t}.
$$

Then, after a change of variables $e^\phi=\rho$, the first equation in \eqref{einsteq} becomes
\begin{equation}\label{liouvillesing}
\Delta\log{\sqrt{\rho}}+ \Lambda \rho=-\sum_i m_i\delta^{(2)}(\vec{x}-\vec{x}_i),
\end{equation}
the well-known Liouville equation with point sources.  From the equation it follows that, outside the sources, the metric has constant curvature $\Lambda$.
From eqns.\ \eqref{metricansatz} and \eqref{voldef}, we find that at time $t$, the universe has volume
$$
\vol(t) = \left(\int\!\rho \, d^2x \right) \cdot \cosh^2 (\sqrt{\Lambda} t),
$$
that is, we have a bouncing solution with initial volume 
$$
\vol(t=0)=\int\!\rho \, d^2x.
$$
We note that the normalized volume density \eqref{densdef} now reads
$$
\tilde{\rho}=\frac{\rho}{\int\!\rho \, d^2x}.
$$
Observe how any dependence on time has disappeared.
 
We now specialize to the case where $X^2=S^2$ is a sphere.

\subsubsection{The sphere}
In the absence of matter ($m_i=0$), we find a unique solution to the Liouville equation \eqref{liouvillesing}, corresponding to the round metric of radius $1/\sqrt{\Lambda}$ (initial volume $4 \pi / \Lambda$), which we write as
$$
ds^2_1=\rho_1(dx^2+dy^2),
$$
where
$$
\rho_1=\frac{4}{\Lambda}\frac{1}{(1+|z|^2)^2}\quad(z\equiv x+y i).
$$

The introduction of point sources gives rise to different, non-round metrics on $S^2$. For example, if
$$
f(z)=\frac{P(z)}{Q(z)}
$$
is a rational function of the complex variable $z=x+iy$, then 
$$
\rho_f=\frac{4}{\Lambda} \frac{|f'(z)|^2}{(1+|f(z)|^2)^2}
$$
gives a solution to equation (\ref{liouvillesing}), for a suitable choice of particle positions and masses.\footnote{These solutions were already studied in connection with the Jackiw-Pi model, for example by Horv\'{a}thy and Y\'{e}ra \cite{HY}.} It follows from \cite{HY} that in this case, the initial volume is
\begin{equation} \vol(t=0) = \frac{4 \pi }{\Lambda} \cdot \mathrm{ord}_\infty(f), \end{equation} 
where $\mbox{ord}_\infty(f)$ is the number of poles of $f(z)$. 

For $\deg(f) \neq 0, 1$, this solution has a different character from the everywhere constant curvature solution $\rho_1$ (which arises for $f(z)=z$). Indeed, at the zeros of
$$
P'(z)Q(z)-P(z)Q'(z),
$$
the metric $\rho_f(dx^2+dy^2)$ has a conical singularity.

\begin{figure*}[h]
\centering
\subfigure[Shown is $\rho_1$, that is, the solution of the Liouville equation corresponding to zero particles ($f(z)=1/z$).]{
\includegraphics[width=6.5cm]{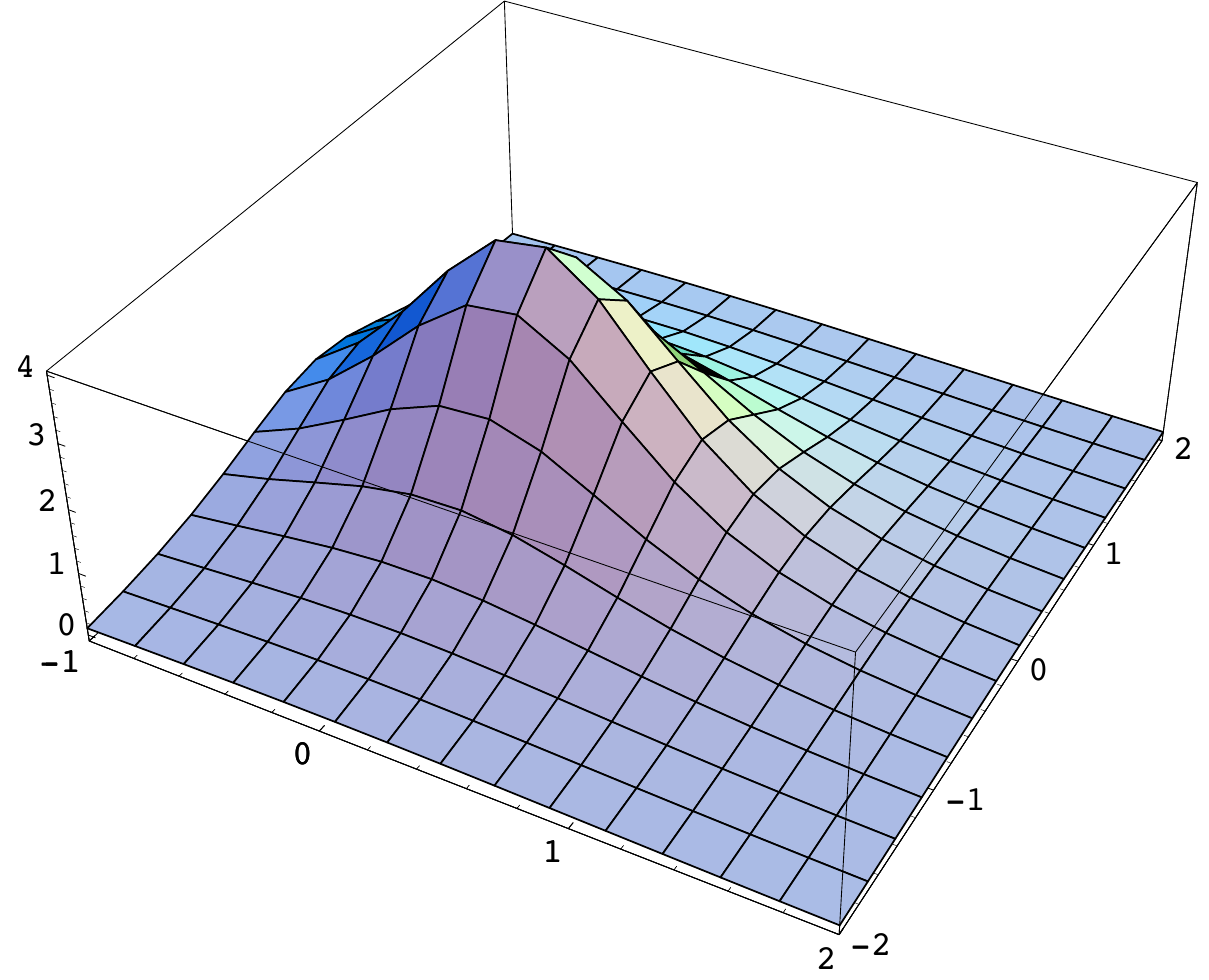}
\label{fig:z}
}
\subfigure[Solution of the Liouville equation associated with the function $f(z)=1/z+1/(z-1)$.]{
\includegraphics[width=6.5cm]{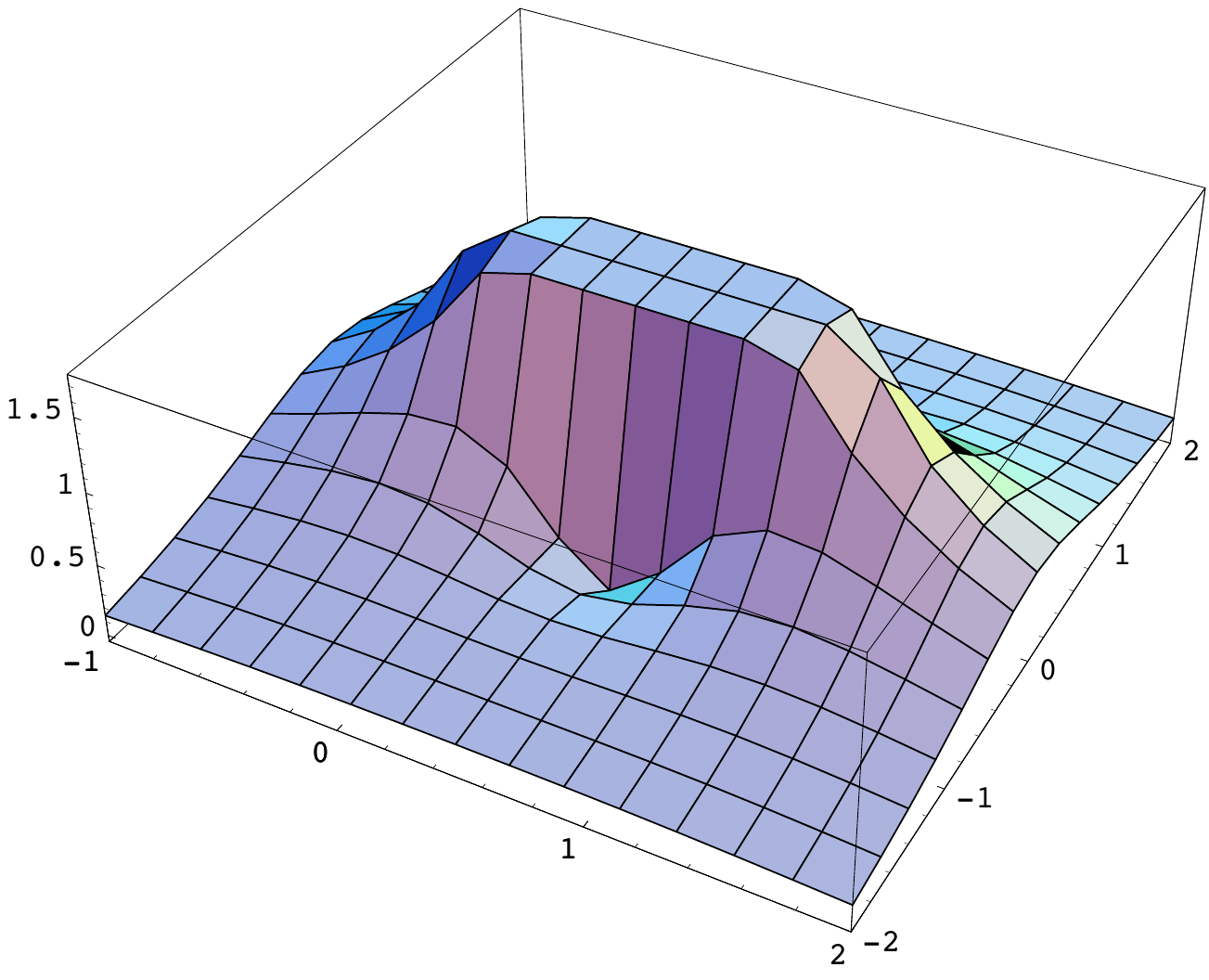}
\label{fig:z-1}
}
\caption{Two solutions of the Liouville equation for $X^2=S^2$.}\label{s2liou}
\end{figure*}

Figure \ref{s2liou} shows a plot of $\rho_f$ where $f(z)=1/z+1/(z-1)$, and also of $\rho_1$.

The striking feature of the relative entropy of $\tilde{\rho}_f$ w.r.t.\ to $\rho_1$, i.e.\
$$
\E(\tilde{\rho}_f\|\tilde{\rho}_1),
$$
is that it is independent of our isothermal coordinates and is sensitive to the lumps we see in the plots.

In the example of a radially symmetric solution $\rho_n=\rho_{z^n}$ corresponding to the power $f(z)=z^n$, it is not too difficult to see that
$$
\E(\tilde{\rho}_n\|\tilde{\rho}_1) = \int\limits_{\R^2} \tilde{\rho}_n \log{\left(\frac{\tilde{\rho}_n}{\tilde{\rho}_1}\right)}\,d^2x=
\log{n}+2(J_n-1),
$$
where
$$
J_n=\int\limits_0^\infty\frac{dx}{(1+x)(1+x^n)}\to\log{2}\quad(n\to\infty).
$$

\begin{figure}[h]
\begin{center}
\includegraphics[width=8cm]{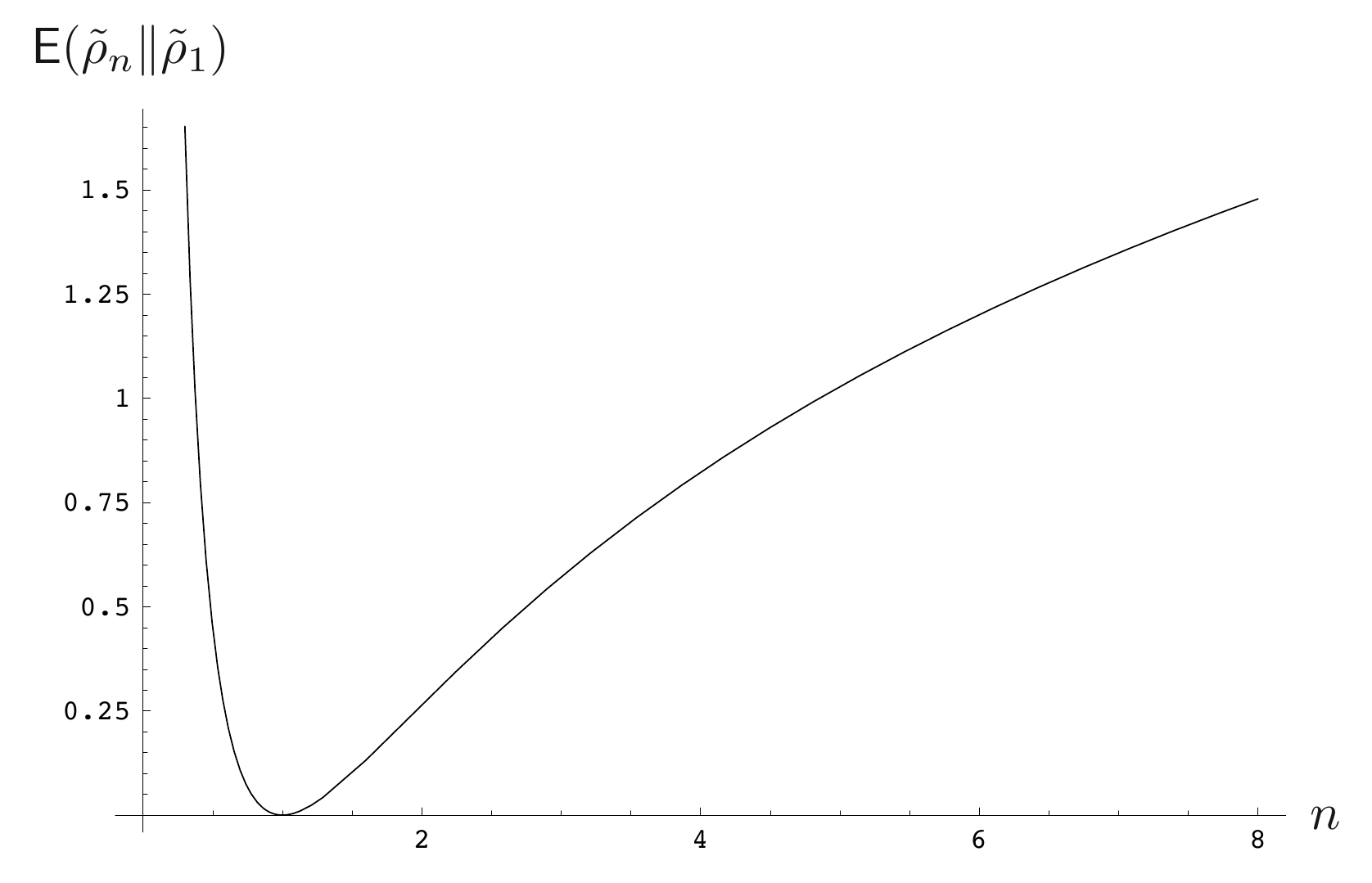}
\caption{$\E(\tilde{\rho}_n\|\tilde{\rho}_1)$ for $X^2=S^2$.}
\label{sphere}
\end{center}
\end{figure}

Physically, the solution $\rho_n$ arises when there are $n-1$ particles of mass $m=-2\pi$ at the north pole of $S^2$ and $n-1$ particles of the same kind at the south pole.\footnote{The appearance of negative masses need not disturb us here, as long as we are talking about a toy model.} Thus, here the relative volume entropy encodes the number of particles $n$. Alternatively, we can think of two particles of mass $m=-2\pi(n-1)$ each, one sitting at the north pole, the other at the south pole, and in this interpretation, $n$ need not be an integer. Figure \ref{sphere} shows a plot of $\E(\tilde{\rho}_n\|\tilde{\rho}_1)$ for $X^2=S^2$.

\subsubsection{The torus}
Here, just as for the sphere, we have at our disposal very many explicit solutions of the Liouville equation with distributional sources \cite{Akerblom:2009ev,Akerblom:2010xb}.

We want to emulate our discussion for the sphere, but immediately we are faced with the problem that on the torus with $\Lambda>0$ there does not exist a zero particle solution (this follows from the Gauss-Bonnet theorem for generalized Riemannian surfaces, compare \cite{Akerblom:2010xb}). The least number of particles allowed is equal to one. The solution with precisely one particle of mass $m=-2\pi$ located at the point $(0,0)$ on the torus was discovered by Olesen \cite{Olesen:1991df,Olesen:1991dg} and is given by
$$
\rho_\mathcal{O}=\frac{4}{\Lambda}\frac{|\wp'|^2|e_1|^2}{(|e_1|^2+|\wp|^2)^2}.
$$
Here,
$$
\wp(z)\equiv\wp_{2,2i}(z)
$$
is the Weierstrass pe-function associated with the lattice $2\Z+2i\Z$, and $e_1=\wp(1)$.

\begin{figure*}[ht]
\centering
\subfigure[Shown here is Olesen's periodic solution of the Liouville equation $\rho_1=\rho_\mathcal{O}$, corresponding to a single particle at $(0,0)$.]{
\includegraphics[width=6.5cm]{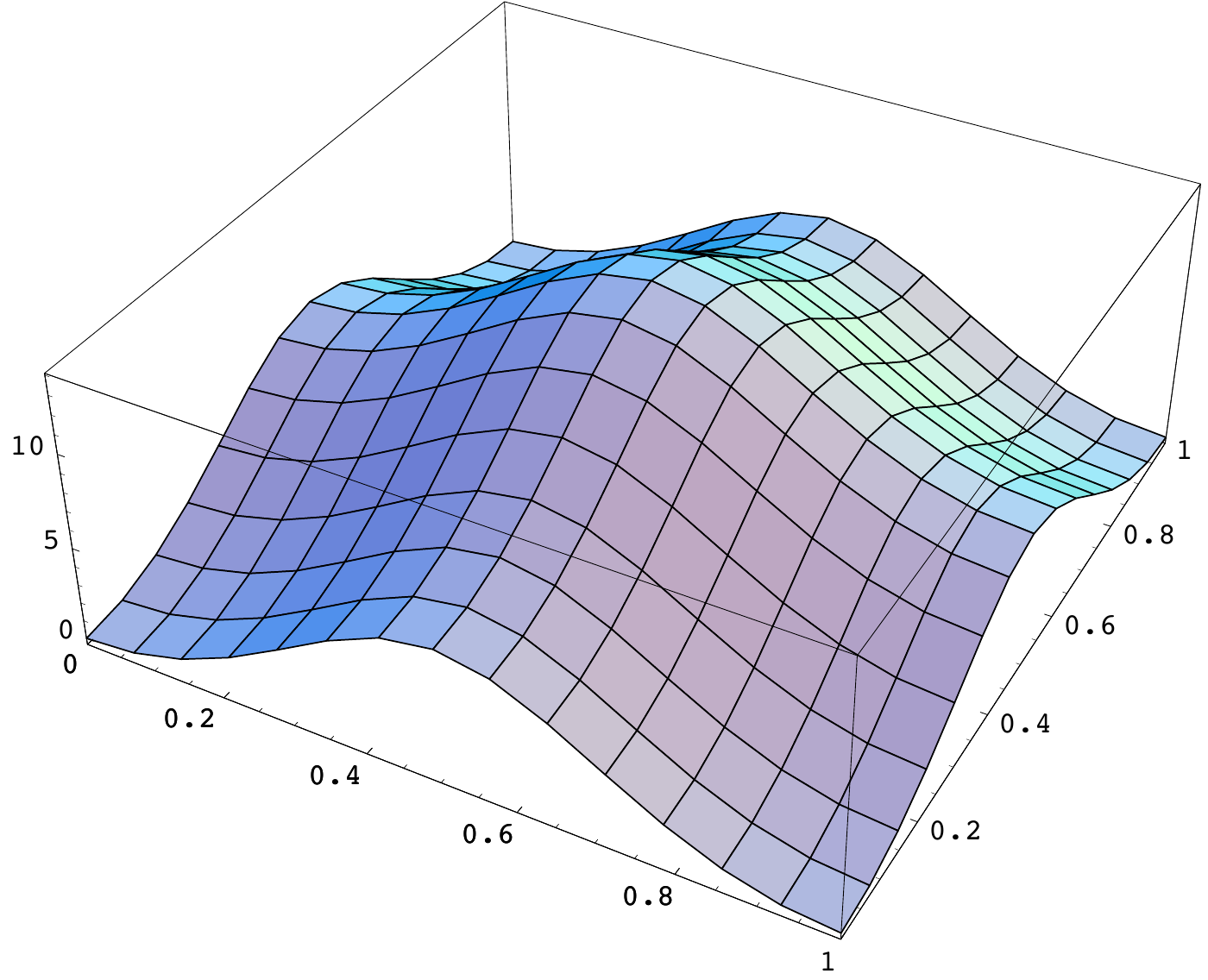}
}
\subfigure[Shown here is $\rho_2$, a solution of the Liouville equation corresponding to one particle at $(0,0)$ and another one at $(0,1/2)$.]{
\includegraphics[width=6.5cm]{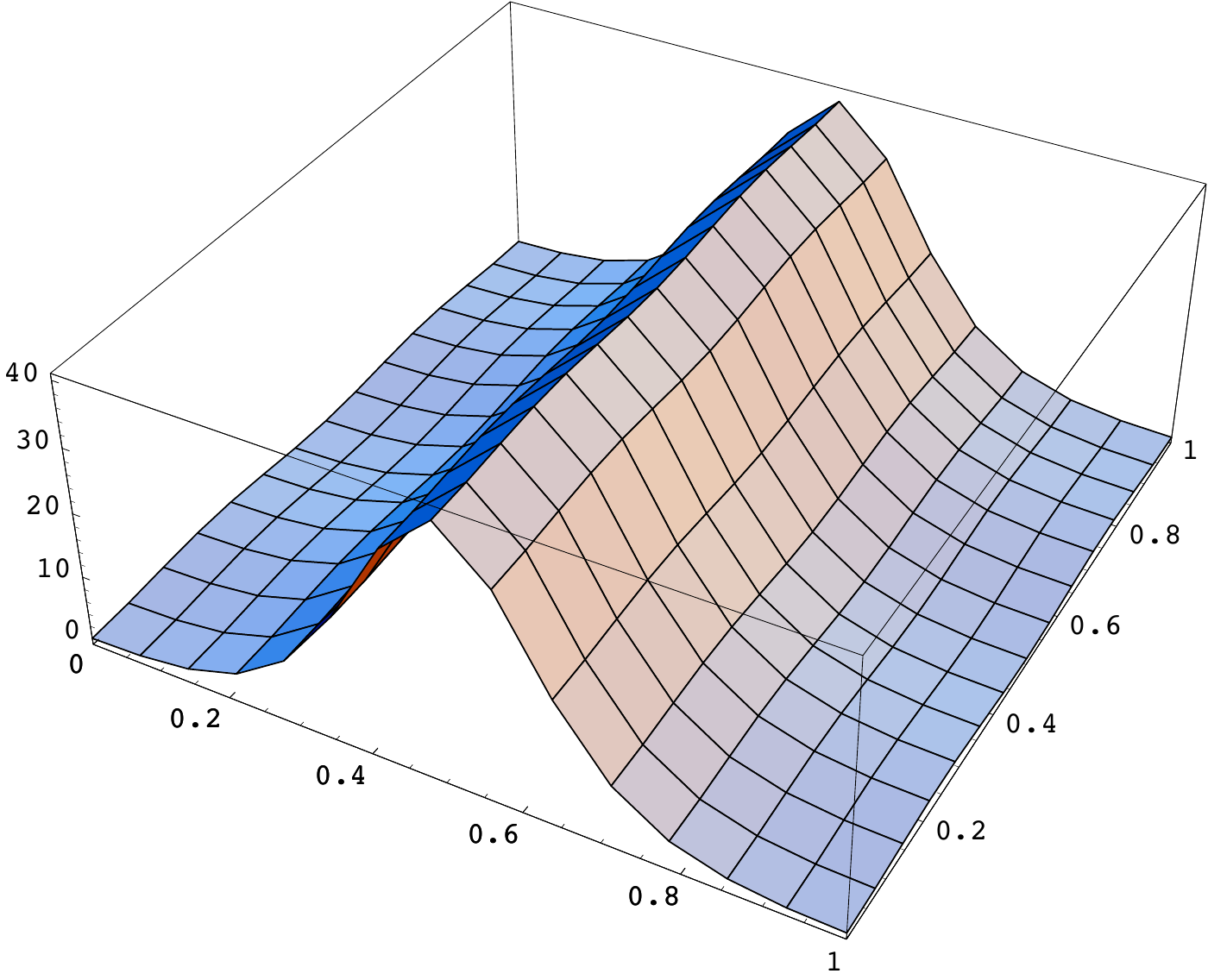}
}
\caption{Two solutions of the Liouville equation for $X^2=T^2$.}\label{olesens}
\end{figure*}

By considering analogous solutions on the lattices 
$$
2\Z+\frac{1}{n}2i\Z\quad (n=1,2,3,\dots),
$$
it is easy to construct $n$ particle solutions of the Liouville equation on the torus (cf.\ \cite{Akerblom:2009ev}, eqns.\ (25), (65), (66)) correspoding to one particle of mass $m=-2\pi$ sitting at each of the points
$$
(0,0), (0,1/n), (0,2/n), \dots
$$
Call these solutions $\rho_n$ (so that $\rho_1=\rho_{\mathcal{O}}$). Examples are shown in Figure \ref{olesens}.

\begin{figure}[ht]
\begin{center}
\includegraphics[width=8cm]{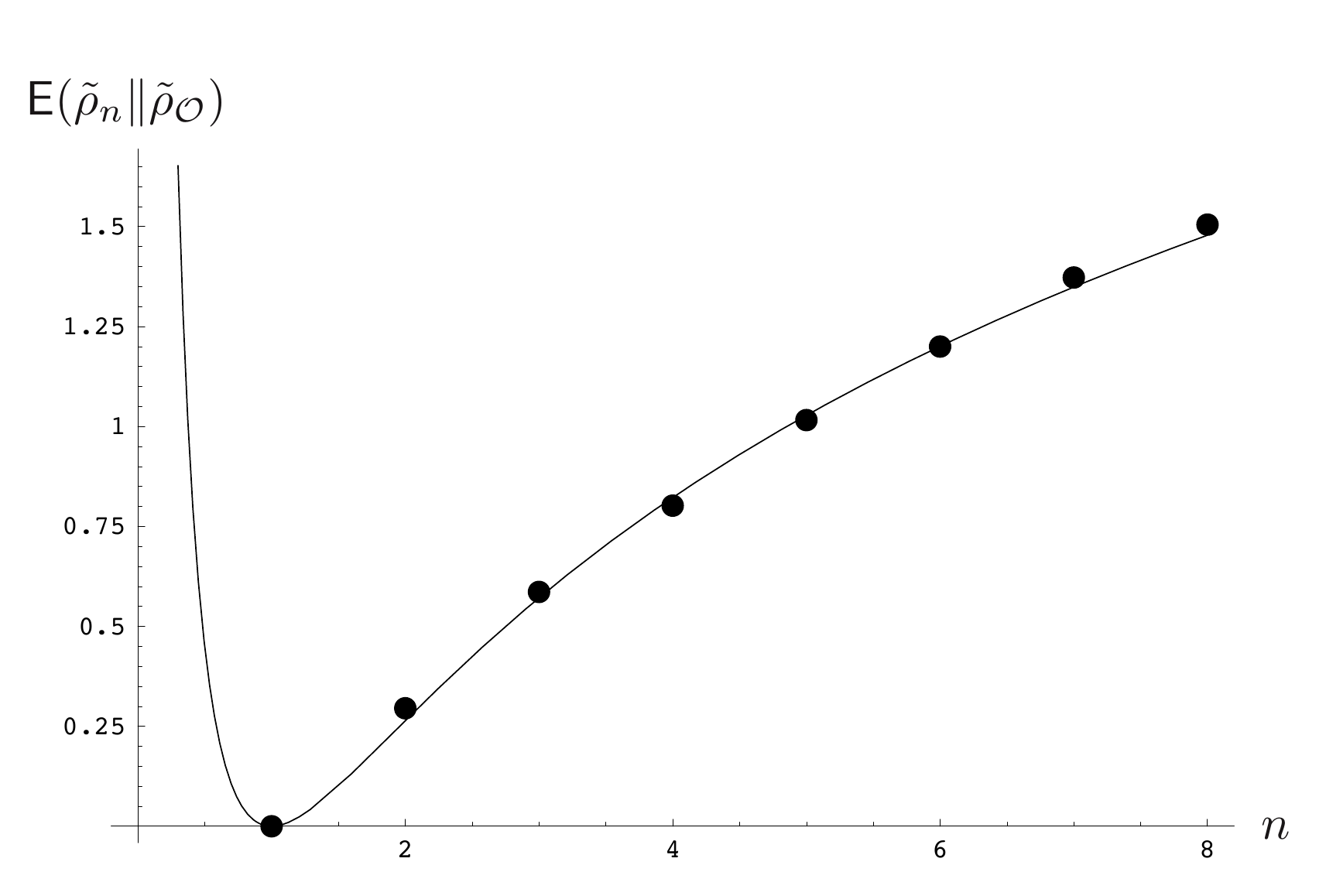}
\caption{$\E(\tilde{\rho}_n\|\tilde{\rho}_\mathcal{O})$ for $X^2=T^2$ (dots) superimposed on the analogous plot for the sphere.}
\label{torus}
\end{center}
\end{figure}

Calculating the relative entropy
$$
\E(\tilde{\rho}_n\|\tilde{\rho}_{\mathcal{O}})
$$
by hand seems like a formidable task. However, numerical computation is simple enough and the outcome is shown in Figure \ref{torus}. Remarkably, the result as a function of $n$ is the same as that for the sphere (within the bounds of numerical accuracy)! This result is the basis for our claim in the Introduction that the relative entropy appears to be quite insensitive to global topology.

\subsection{Gravitational waves}\label{Wave}
Our final example is the relative volume entropy of an exact gravitational wave on a torus w.r.t.\ Minkowski space.\footnote{To the best of our knowledge the class of periodic gravitational waves presented below is new.}

We first exhibit a class of gravitational waves in toroidal space. Let
$$
M=\R_t\times [0,L]^3\quad (L\cong 0).
$$
For the metric, we make the plane wave ansatz
$$
ds^2=-dt^2+f(z-t)^2dx^2+g(z-t)^2dy^2+dz^2,
$$
where $f(u)$ and $g(u)$ are $L$-periodic functions. The vacuum Einstein equations now imply
\begin{equation}\label{ode}
\frac{f''}{f}+\frac{g''}{g}=0.
\end{equation}
We can write down $L$-periodic solutions to this equation in terms of Mathieu functions\footnote{Compare Whittaker and Watson \cite{Whittaker} or \cite{Weiss}.} as follows.

Let $q_0>0$ be a solution to the equation
$$
a_n(q_0)=-b_m(-q_0)\quad(m,n\in\N^*),
$$
where the $a$s and $b$s are the characteristic values of even, respectively odd Mathieu functions.

Then, using the differential equation satisfied by Mathieu functions, it is easy to see that the following is a doubly infinite system of real $L$-periodic solutions to \eqref{ode}:
$$
g(u)=C\Big(a_m(q_0), q_0, \frac{2\pi u}{L}\Big)\quad\text{(Mathieu cosine function)},
$$
and
$$
f(u)=S\Big(-b_n(-q_0),-q_0,\frac{2\pi u}{L}\Big)\quad\text{(Mathieu sine function)}.
$$
Thus, our solutions depend on two parameters $m,n\in\N^*$.

\begin{figure}[ht]
\begin{center}
\includegraphics[width=8cm]{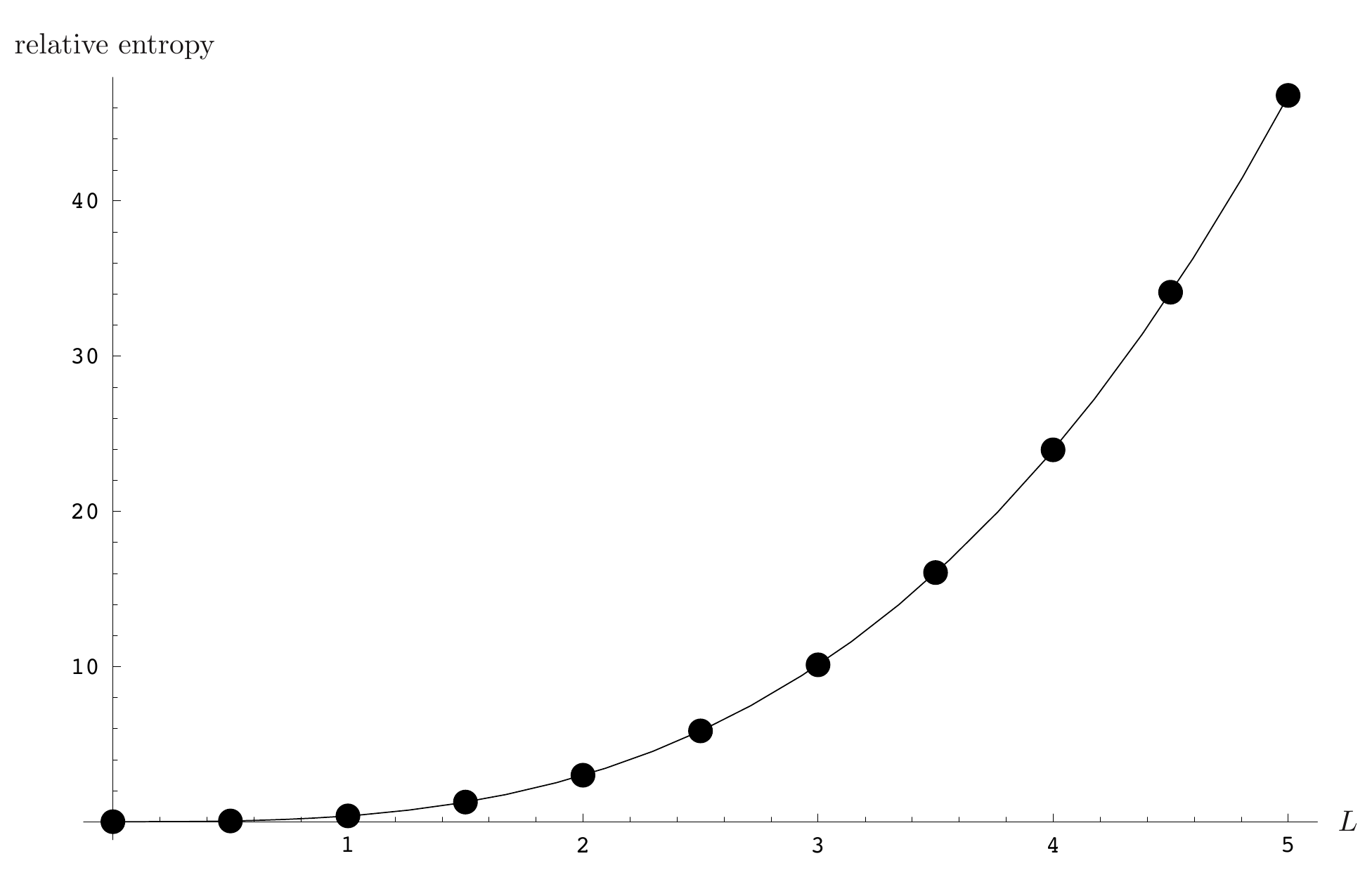}
\caption{Relative volume entropy w.r.t.\ Minkowski space of a gravitational wave with $m=1$ and $n=2$ as a function of the period-length $L$ (dots), superimposed on a plot of the function $x\mapsto 0.374\, x^3$.}
\label{wave}
\end{center}
\end{figure}

We have studied the relative volume entropy of a sample of these solutions w.r.t. Minkowski space as functions of the period-length $L$. We have found that invariably the relative volume entropy scales like $L^3$. Figure \ref{wave} is one example of this behavior.

\section{Discussion}

In this paper, we have introduced the relative entropy of volume, an information theoretical measure in classical gravity, measuring part of the information content of one solution of the Einstein equation relative to another. We have studied in some examples how it detects the deviation from homogeneity as caused by gravitational radiation and matter. In a technical direction, it would be interesting to extend the theory beyond the case of compact spacelike slices and metrics with a preferred time direction.

It would also be interesting to study the relative volume entropy in the presence of a non-trivial Weyl tensor, for example in the light of Penrose's proposal of non-activation of gravitational degrees of freedom (encoded in the Weyl tensor) at the Big Bang \cite{Penrose}.

In addition, believe that relative volume entropy could play a role in the theory of cosmic structures (compare \cite{Springel:2006vs}). Because the gravitational fields involved in the formation of structures in the universe are quite weak, it is sufficient to ignore the backreaction on spacetime geometry in the pertinent numerical simulations. Notwithstanding this appropriate practice, we believe that minimizing the relative entropy of a putative analytical model of the resulting (backreacted) spacetime w.r.t.\ to numerical data would provide a way of fixing parameters in the analytical model (applying the ``principle of mean discrimination of information'' due to Kullback).  

The natural approach to entropy in classical and quantum gravity often is the von Neumann entropy (see e.g.\ \cite{Brandenberger:1992jh,VanRaamsdonk:2010pw}), and it would certainly be very interesting to study its relation to the present relative entropy. As an example of this, in loop quantum gravity, volume is a (trace class) operator $\mathbf V$ (compare \cite{Rovelli,Loll:1995wt,Lewandowski:1996gk}), so it makes sense to consider its von Neumann entropy $-\mathrm{Tr}(\mathbf V \log \mathbf V)$ (as a regularized value). Observe that in this case, there is no need for a \emph{relative} version of entropy (or one may consider it as a relative entropy over empty spacetime). Does this ``loop volume'' entropy have a relation to our relative volume entropy (over a suitable background)?

\paragraph{Acknowledgments.} It is our pleasure to thank Volker Springel for a briefing on cosmic structure formation and Jan-Willem van Holten for valuable discussions about gravitational waves. In addition, we would like to thank Jan-Willem van Holten and Renate Loll for helpful comments on the manuscript.

\bibliographystyle{utphys}

\bibliography{gravity}

\end{document}